\documentclass{aa}

\usepackage{graphicx}
\usepackage{subfigure}

\usepackage{txfonts}
\usepackage{hyperref}
\hypersetup{
    colorlinks = true,
    citecolor ={blue}
}

\begin{document} 

   \title{Discerning internal conditions of pulsating hot subdwarf B stars
   }
   \subtitle{Modelling multiple trapped modes in KIC\,10001893}

   \author{Hamed Ghasemi\inst{1} \thanks{
    \email{hamed.ghas@gmail.com}},
        Murat Uzundag\inst{2},
        Cole Johnston\inst{3,4,2,5},
        Conny Aerts \inst{2,5,6}
        }
   \institute{Independent Scholar, 9000, Ghent, Belgium
   \and
    Institute of Astronomy, KU Leuven,
    Celestijnenlaan 200D, 3001, Leuven, Belgium
    \and
    Astrophysics group, Department of Physics, University of Surrey,
    Guildford, GU2 7XH, United Kingdom
    \and
    Max-Planck-Institut für Astrophysik, Karl-Schwarzschild-Straße 1, 85741 Garching bei München,Germany 
    \and
    Radboud University Nijmegen, Department of Astrophysics, IMAPP, P.O. Box 9010, 6500 GL Nijmegen, The Netherlands
    \and
    Max Planck Institute for Astronomy, Koenigstuhl 17, 69117 Heidelberg, Germany
    \\
    }

\date{Received 19 February 2025 / Accepted 13 May 2025}

    \titlerunning{}
    \authorrunning{H. Ghasemi et al.}

 
  \abstract
   {The frequencies of gravity mode oscillations are determined by the chemical, thermal, and structural properties of stellar interiors, facilitating the study of internal mixing mechanisms in stars. We investigate the impact of discontinuities in the chemical composition induced by the formation of an adiabatic semiconvection region during the core helium (He)-burning phase of evolution of hot subdwarf B-type (sdB) stars. }  
   {Our objective is to delve into the progression of convective core evolution by using a numerical approach to model the emergence of a semiconvection zone. We scrutinize the asteroseismic attributes of the evolutionary stages and assess the core He-burning phase by evaluating the parameter
   linked to the average interval between the deep trapped modes in both sdB evolutionary models and the observations of KIC 10001893.}
   {We perform evolutionary and asteroseismic analyses of sdB stars using {\sc mesa} and {\sc gyre} to examine the properties of the semiconvection region. Additionally, we compute parameters related to gravity-mode period spacings and the interval between deep trapped modes to characterize the core helium-burning phase at different stages of sdB evolution.}
   {Our study illustrates the evolution of the convective core in sdB stars by using a numerical scheme in {\sc mesa} to model the development of the semiconvection zone. We address the challenges of relying solely on average interval between oscillation mode periods with consecutive radial orders to identify the core He-burning stage. To enhance identification, we propose a new parameter representing the average interval between deep trapped modes during some of the stages of sdB evolutionary models. Additionally, our results show that integrating convective penetration with convective premixing improves our models and yields comparable outcomes without the need for additional model parameters.}
   {Our results can advance the development of detailed evolutionary models for sdB stars by refining internal mixing schemes, enhancing the accuracy of pulsation predictions, and improving alignment with observational data.}


   \maketitle
%

\section{Introduction}

Hot subdwarf B stars, known as sdBs, are believed to be core-helium (He) burning stars characterized by an exceedingly thin hydrogen (H) envelope, with an envelope mass less than 0.02 times the mass of the Sun. Their average mass aligns closely with the mass threshold at which the core-He flash occurs, approximately around 0.47 times the mass of the Sun \citep{2012A&A...539A..12F}. These sdB stars represent evolved, compact entities with surface gravities ($\log{g}$) ranging from 5.2 to 6.2 dex and effective temperatures ($T_{\rm eff}$) spanning from 20\,000 to 40\,000 K, and radii ranging from 0.10 $R_{\odot}$ to 0.30 $R_{\odot}$ \citep{Heber2016}. They occupy the intermediate phase between the main sequence and the cooling stage of white dwarfs, known as the extreme horizontal branch (EHB) stars \citep[see][for a detailed review]{Heber2016}.

The evolutionary trajectory of sdB stars requires substantial mass loss, primarily driven by binary interactions during the late stages of the red giant phase \citep{paczynski1976,han2002, han2003, Pelisoli-2021}. This phase leads to the removal of nearly all of the H-rich envelope, resulting in a core that burns He but possesses an envelope too thin to sustain H-shell burning.
Over a period of roughly $10^{8}$ years, sdB stars continue to burn He in their cores. Upon the exhaustion of He in their core, they progress into a phase where He is burned in a shell surrounding a core composed of carbon and oxygen (C/O), evolving into subdwarf O (sdO) stars. Ultimately, these stars end their lifecycle as white dwarfs \citep{dorman1993}. This exotic evolutionary pathway affords us the opportunity to directly observe and scrutinize the properties of the mixed He-C/O cores of low- and intermediate-mass stars, which are otherwise obscured by the thick hydrogen envelope.  

\citet{kilkenny1997} were the first to discover rapid pulsations in a subgroup of hot sdB stars, now termed V361 Hya stars, commonly known as short-period sdB variable stars. These stars exhibit multiple pulsation periods ranging from 60 s to 800 s, corresponding to low-degree (\(\ell \lesssim 3\)), low-order (0 \(\leq  n \lesssim 20\)) pressure (p)-modes within this frequency range. The excitation of these modes is attributed to a classical $\kappa$-mechanism, primarily driven by the accumulation of iron group elements, particularly iron itself, in a region known as the $Z$-bump. This mechanism was proposed by \citet{charpinet1996, charpinet1997} showing that radiative levitation enhances the concentrations of iron group elements, a prerequisite for activating the pulsational modes.
The p-mode sdB pulsators typically populate a temperature range of 28\,000 K to 35\,000 K, with surface gravities between $\log g$ 5 to 6 dex. 
Following this discovery, another class of sdB pulsators, termed V1093 Her stars, displaying long-period pulsations, was identified by \citet{green2003}. These stars exhibit brightness variations with periods extending to a few hours.
The oscillation frequencies observed in these pulsators are associated with gravity (g)-modes characterized by low-degree ($\ell <$ 3) and medium- to high-order (10 $< n <$ 60), driven by the same $\kappa$-mechanism due to the accumulation of iron group elements, as outlined by \citep{fontaine2003,Jeffery-2006a,Jeffery-2006b,Jeffery-2007, Charpinet-2011}. In contrast to p-mode sdB pulsators, g-mode sdB pulsators exhibit somewhat cooler temperatures ranging from 22\,000 K to 30\,000 K, with surface gravities ($\log{g}$) typically falling within 5.0 dex to 5.5 dex.
Within the group of pulsating sdB stars, there is a subset known as "hybrid" sdB pulsators, exhibiting both g- and p-modes concurrently \citep[e.g.][]{schuh2006}. These hybrid sdB pulsators are situated on the borderline between $T_{\rm eff}$ of 28\,000 K and 32\,000 K.
These objects are particularly significant as they offer a unique opportunity for asteroseismic investigations, enabling the study of both the core structure and the outer layers of sdBVs.

The advancements in high precision and high duty cycle photometric monitoring from space missions such as the Kepler mission \citep{borucki2010} and the TESS mission \citep{Ricker2014} have made significant contributions to identifying new candidates of pulsating sdBs. As a result, over 300 pulsating sdBs have been found and documented thus far \citep{Uzundag-2024}.
These missions have not only been effective in identifying new candidates of pulsating sdBs but have also enabled unprecedented asteroseismic measurements.
For instance, seismic tools such as rotational multiplets and asymptotic period spacings have become accessible for pulsating sdBs \citep[see][for a review]{Reed2018, Lynas2021}.

Throughout the horizontal branch (HB) phase, the interaction between layers abundant in C/O and those rich in He causes a change in the chemical composition at the core boundary. The size of the core during the horizontal branch phase can either increase or remain unchanged depending on how the transport or mixing of chemical species near the convective core is modelled in 1D or multi-dimensional approaches, which have been debated at length in the literature \citep{Salaris-2017, Paxton_2019, Herwig-2023, Blouin-2024}. Various mixing mechanisms have been proposed, such as semiconvection, convective penetration, and convective overshooting \citep{Castellani_1985, Mowlavi_1994, Bossini-2015, Schindler-2015, Constantino-2015, Schindler-2017, Xiong-2017, Li-2018, Guo-2018, Johnston-2024}. In addition to the chemical mixing outside of the core, the method for determining the location of the convective boundary introduces uncertainty into evolutionary calculations. Some numerical approaches to modelling mixing processes in stellar evolution aim to represent the emergence of adiabatic semiconvection regions that extend beyond traditional convection.\citep{Paxton_2019,castellani1985, Mowlavi_1994}. Semiconvection persists during the helium-burning phase, with the semiconvective region achieving convective neutrality both at the edge of the core and throughout its extent. Maintaining the balance of He, C, and O within this semiconvective zone is crucial as it regulates the necessary adjustment in the radiative gradient of this region. The development of adiabatic semiconvection beyond the convective core presents a promising alternative to overshooting approaches. This method effectively circumvents issues such as convective shell formation. The precise effects of various mixing scenarios and their corresponding numerical methods remain a topic of debate. However, current research is revealing promising new directions. Two promising approaches involve modelling the R2-ratio, which compares the number of stars observed during the early asymptotic giant branch (eAGB) phase to those in the horizontal branch (HB) phase \citep{Constantino-2016, Constantino-2017, Ghasemi-2017} and asteroseismology. 

The non-radial oscillations observed in pulsating sdBs offer a unique way to probe the internal properties of these stars if their pulsation geometry can be determined \citep{Aerts2010}. Space observations have facilitated the identification of oscillation modes via pattern recognition, a capability that has hardly been achievable with ground-based data. This development has had a profound impact on theoretical models, particularly in areas where physics related to chemical mixing (especially diffusion, overshooting and semiconvection) are poorly constrained. Exploring pulsating compact stars, particularly sdB stars, offers a unique approach to understanding mixing mechanisms in the vicinity of their core and envelope through asteroseismology \citep{Charpinet-2002a, Charpinet-2002b, Charpinet-2011, VanGrootel-2008, VanGrootel-2010a, VanGrootel-2010b, Hu-2007, Hu-2008, Hu-2009, Hu-2010, Hu-2011, Bloemen-2014}. In particular, the power of modelling pulsations has already been shown to allow for the characterization of the chemical profile and mixing mechanisms in sdB stars \citep{Constantino-2015, Ghasemi-2017,Charpinet-2019, Ostrowski-2021}. 

Asymptotic period sequences, particularly for dipole ($\ell=1$) and quadrupole ($\ell=2$) modes of consecutive radial order, have proven effective for analyzing the g-mode pulsations in the observed sdBV stars with over 60\% of the periodicities being attributed to these modes \citep[e.g.,][]{reed2011, Uzundag2021}. However, the sdB stars are characterized by layers of varying composition, some of which result from mixing mechanisms near their cores. This complexity complicates the straightforward application of asymptotic approximations typically used for homogeneous stars. These compositional discontinuities cause deviations from sequences expected for homogeneous stars. Such disruptions have been observed in several pulsating sdB stars. Furthermore, as these compositional discontinuities intensify in the transition zones within the stars due to the mixing process, certain modes may become deeply trapped near the convective core \citep{Constantino-2015, Ghasemi-2017,Ostrowski-2021,Guyot-2025}. This phenomenon has also been identified in a few pulsating sdB stars observed by Kepler \citep{ostensen2014, Uzundag2017}. 

Throughout the nominal Kepler mission, a total of 18 pulsating subdwarf B stars were observed in short-cadence mode. Among these stars, the majority (16) exhibit long-period g-mode pulsations, with only two showing short-period p-mode pulsations. Furthermore, three previously identified sdB stars within the open cluster NGC 6791 were discovered to pulsate \citep{Reed2018}. 
In this study, we focused on modelling a pulsating sdB star, namely KIC\,10001893, one of the field sdB stars, which underwent extensive monitoring by the Kepler spacecraft. The spectroscopic atmospheric parameters of KIC\,10001893, as provided by \citet{ostensen2011}, including $T_{\rm eff} = 26\,700 \pm 300$ K, $\log{g} = 5.30 \pm 0.04$, and $\log(\rm N_{He}/\rm N_{H}) = -2.09 \pm 0.1$, are strongly indicative of alignment with the g-mode instability strip \citep{2014A&A...570A.130S}. Frequency analysis of KIC\,10001893 unveiled 110 oscillation peaks. A comprehensive seismic analysis of KIC\,10001893 was conducted by \citet{Uzundag2017}, revealing 32 $\ell = 1$ and 18  $\ell = 2$ modes. The authors also discovered the presence of nearly complete sequences of consecutive radial orders for both $\ell = 1$ and $\ell = 2$ modes, which included three deep trapped modes. These trapped modes were clearly visible in the reduced-period diagram, which displayed an almost perfect alignment of the two sequences. The locations of these deep trapped modes, along with the spacing between them, provide invaluable information for examining the stellar interior and evaluating various mixing models in the vicinity of the core boundary. 

This paper is organised as follows. Section~\ref{sec:2} explains the models for convective core evolution in sdB stars. It covers the sign-change algorithm, which utilizes overshooting and detects convective boundaries using either the Schwarzschild criterion or the Ledoux criterion. Additionally, this section explains the predictive mixing (PM) numerical model, which extends convective regions until the boundaries meet convectivity conditions on the convective side of the core within a single time step. Finally, Section~\ref{sec:2} introduces the convective premixing (CPM) scheme, which employs a numerical approach to model the emergence of the adiabatic semiconvection zone during the core helium (He)-burning phase. This section also describes the evolution and formation of sdB stars, as well as the evolution of the convective core incorporating the CPM scheme. Section~\ref{sec:3} describes the asteroseismic properties of sdB models employing the CPM scheme. In Section~\ref{sec:4}, we investigate the asteroseismic properties of sdB models with time-dependent convective penetration in conjunction with the CPM scheme. Finally, in Section~\ref{sec:5}, we discuss our findings and potential future directions. The CPM scheme and the convective penetration implementation in {\sc mesa} models for sdB stars are publicly available at github\footnote{https://github.com/Hamedghas/Hot-subdwarf-B-type}.

\section{Models of Convective Cores in Evolutionary Models of sdB Stars}
\label{sec:2}
We calculate stellar structure and evolution models using the 1D Modules for Experiments in Stellar Astrophysics code {\sc mesa} \citep{Paxton_2011,Paxton_2013,Paxton_2015,Paxton_2018,Paxton_2019,Jermyn_2023}. Determining the location of convective boundaries during both the core H- and He-burning phases is a complex subject. \citet{Paxton_2019} discuss three numerical methods available in {\sc mesa}: the sign-change algorithm, predictive mixing (PM), and convective premixing (CPM) scheme (see Figure~\ref{fig1}). The details of our computational models of sdB stars using {\sc mesa} are discussed in Appendix ~\ref{app:a}. The simple sign-change method for determining convective boundaries involves detecting sign changes in the variable $y$ using either the Schwarzschild criterion ($y = \nabla_{\text{rad}} - \nabla_{\text{ad}}$) or the Ledoux criterion ($y = \nabla_{\text{rad}} - \nabla_{\text{L}}$). Here, $\nabla_{\text{rad}}$, $\nabla_{\text{ad}}$, and $\nabla_{\text{L}}$ represent the radiative, adiabatic, and Ledoux temperature gradients, respectively. However, this approach may fail in circumstances with composition discontinuities, particularly when the radiative gradient $\nabla_{\text{rad}}$ exceeds the adiabatic gradient $\nabla_{\text{ad}}$ on the convective side of the core \citep{Gabriel_2014}. In the outer region of the convective core, overshooting leads to increased transfer of C and O into the radiative part, consequently increasing opacities. Additionally, following Kramer's law, temperature decreases toward the outer part of the convective core, further increasing opacity and $\nabla_{\text{rad}}$ in these outer regions. In the subsequent evolutionary stages, convective regions located below the outer convective layers exhibit lower $\nabla_{\text{rad}}$, transitioning to a radiative state, while convective shells form simultaneously in the outer regions (see Figure~\ref{fig1a}). In the top panel of Figure ~\ref{fig2}), the green plot illustrates the convective shell where $\nabla_{\text{rad}}$ is greater than $\nabla_{\text{ad}}$.

In Figure~\ref{fig1b}, another numerical approach, the PM scheme, improves the sign-change algorithm by extending convective regions until the boundaries satisfy $\nabla_{\text{rad}} = \nabla_{\text{ad}}$ on the convective side within a single time step. This numerical method is modified to prevent the initial splitting of a convection region. During the predictive mixing iterations, if at any point within the convective region $\nabla_{\text{rad}} = \nabla_{\text{ad}}$, the code restricts further growth of that convection region. The PM has demonstrated its effectiveness in achieving the desired objective of preventing the formation of convective shells above the convective core. Nevertheless, as illustrated in the top panel of Figure~\ref{fig2} by the red plot, in this numerical method, the persistence of $\nabla_{\text{rad}} > \nabla_{\text{ad}}$ on the convective side of the convective boundary highlights the necessity for further exploration of mixing models \citep{Gabriel_2014}. 

\begin{figure}[ht]
    \centering
    \subfigure[]{
        \includegraphics[width=0.5\textwidth]{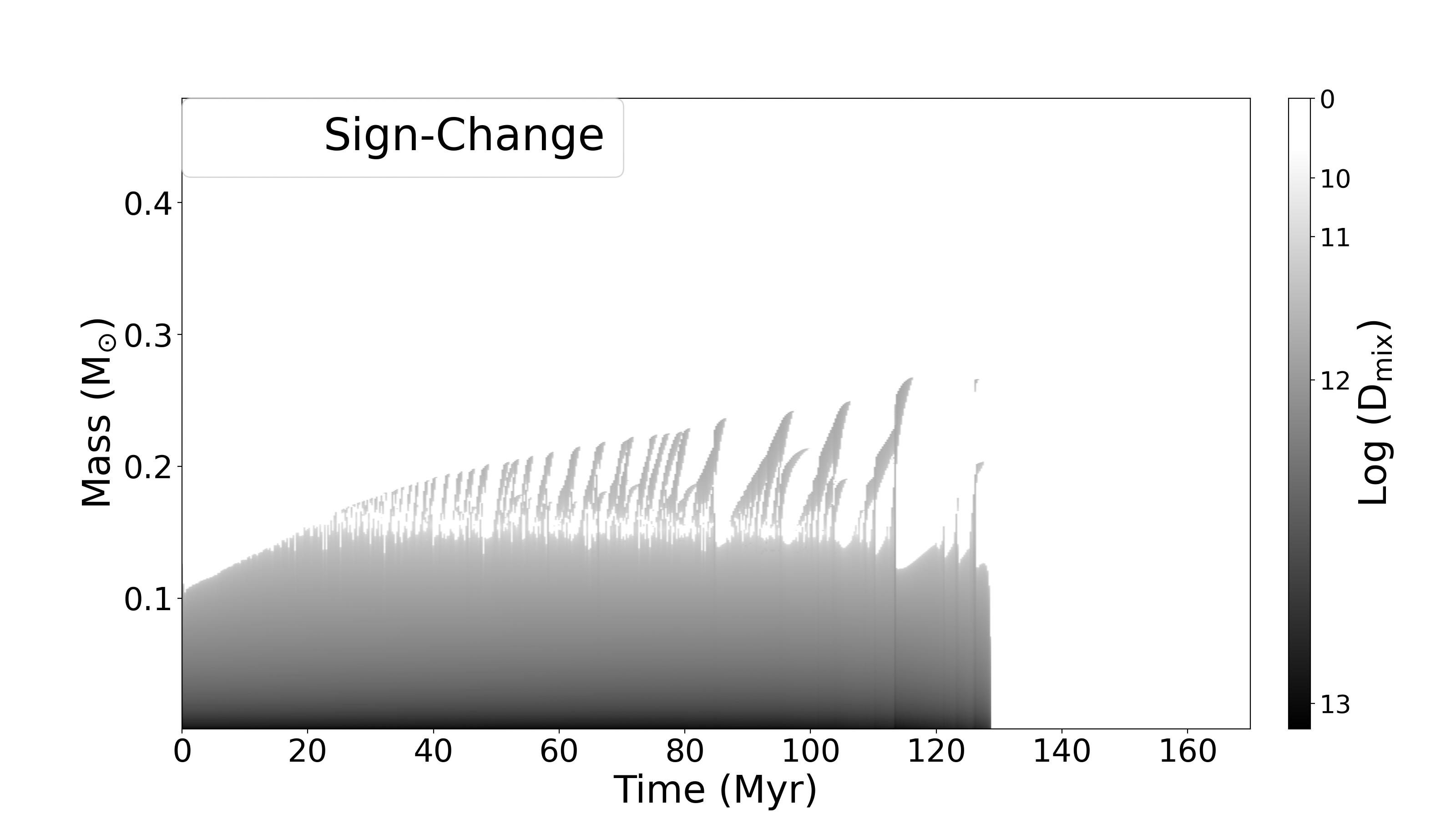}
        \label{fig1a}
    }
    \hfill
    \subfigure[]{
        \includegraphics[width=0.5\textwidth]{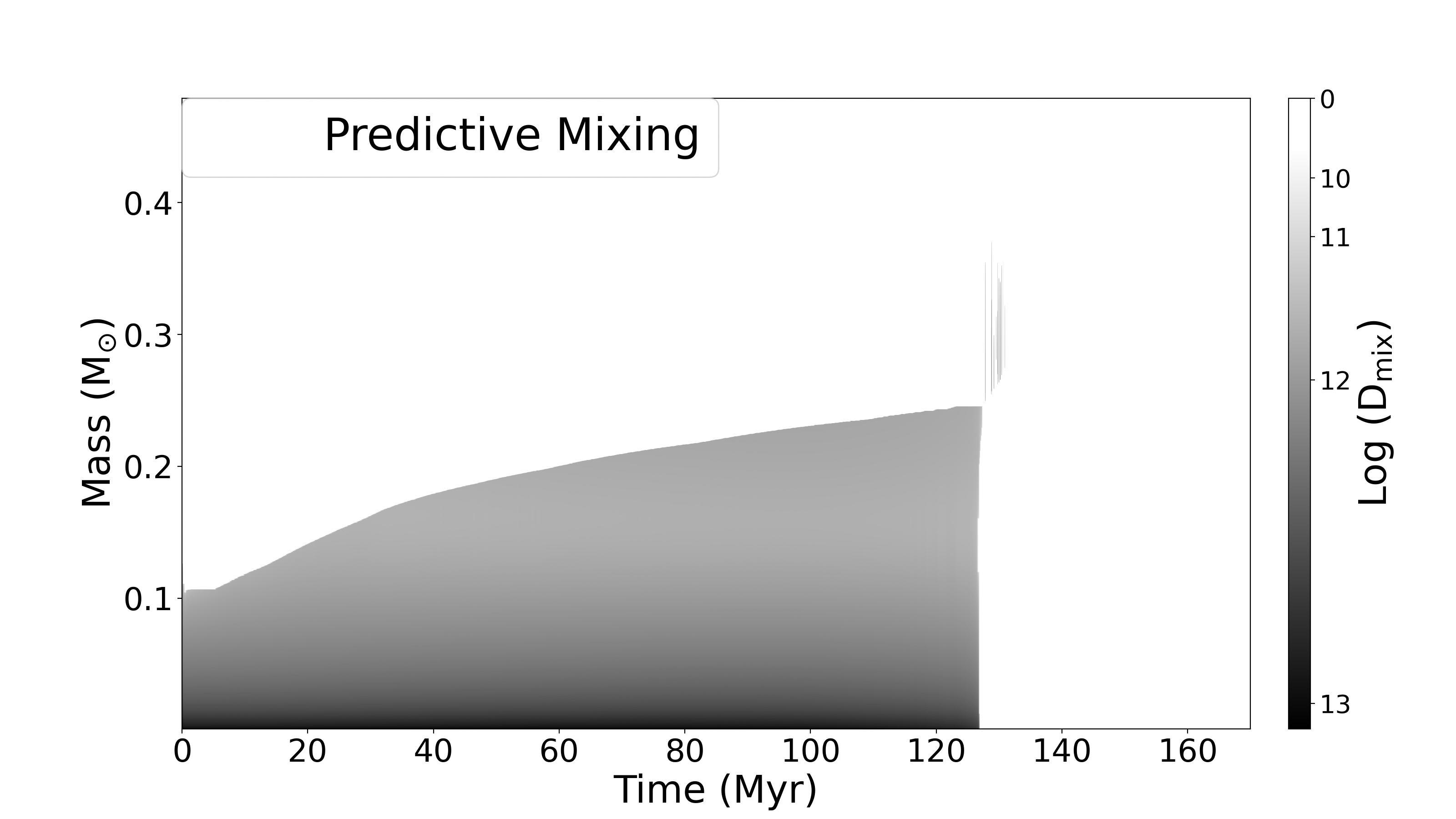}
        \label{fig1b}
    }
    
    \vskip\baselineskip
    \subfigure[]{
        \includegraphics[width=0.5\textwidth]{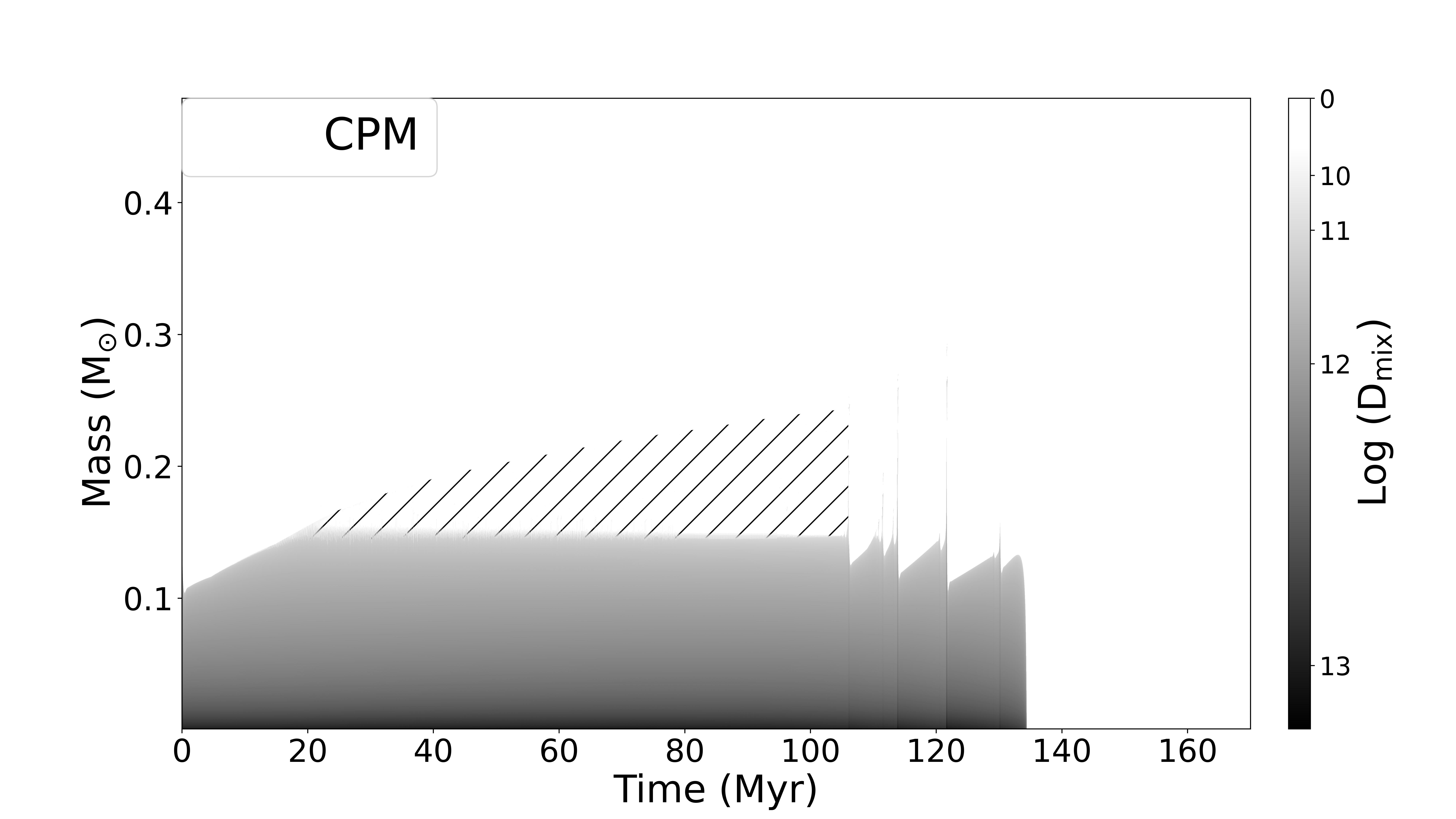}
        \label{fig1c}
    }
    \caption{Schematic representation of the interior structures of an sdB star, displaying the logarithm of the diffusion coefficient for mixing during the core He-burning stage. Different levels of mixing are indicated in the gray-scale bar on the right-hand y-axis.  \textbf{Panel (a)} presents the result of the sign-change numerical method, \textbf{Panel (b)} depicts the result of the predictive mixing, and the lower panel displays the result of the convective premixing scheme. gray color denotes convective core regions, with the hatched area in \textbf{panel (c)} indicating semiconvection in the transitional zone between convective and radiative regions.}
    \label{fig1}
\end{figure}

\begin{figure}[ht]
    \includegraphics[width=0.5\textwidth]{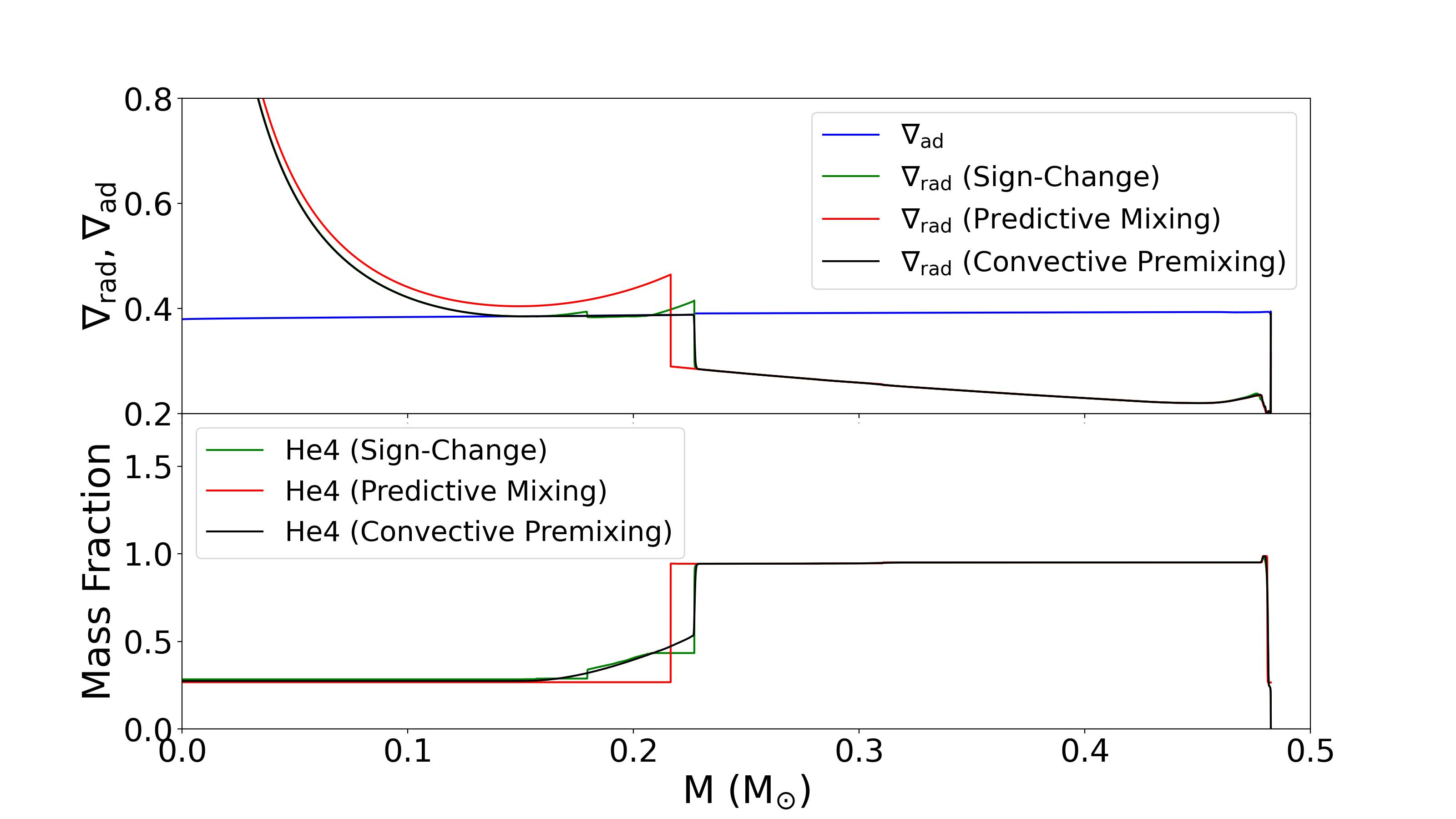}
    \caption{\textbf{Top Panel:} The adiabatic gradient is depicted in blue, alongside radiative gradients for three distinct sdB models, plotted as a function of mass coordinate at 80 Myr. The models are distinguished by the sign-change (green), predictive mixing (red), and convective premixing (black). \textbf{Bottom Panel:} He mass fraction corresponding to each of the three models with temperature gradients in the top panel.}
    \label{fig2}
\end{figure}
As illustrated in Figure~\ref{fig1c}, the CPM scheme provides an alternative numerical technique for modelling mixing beyond convection in stellar evolution, avoiding issues found in previously discussed numerical methods \citep{Paxton_2019}. It is performed at the beginning of each time step before any structural or chemical modifications occur. The CPM extends convective boundaries until the radiative gradient equals the adiabatic gradient on the convective side of the boundary. This stage is often accompanied by the initial appearance of convective shells, which in the CPM scheme gradually merge into the semiconvective zone. The convective shells with shorter mixing timescales approach convective neutrality first. This process continues until all convective shells reach neutrality, causing the breakdown of earlier convective layers and the expansion of the semiconvective zone. A mesh cell above the convective boundary may become convective, while a cell within the convective zone may revert to a radiative state. This process repeats until the outer boundary cells of the convective zone stabilize in radiative states after mixing. Mixing within the mesh cells assumes constant pressure and temperature in each cell, requiring updates to abundances, densities, opacities, and temperature gradients (including radiative, adiabatic, and Ledoux gradients) within the affected cells. Overall, this numerical scheme creates classical semiconvection regions, where the abundance gradient is adjusted to maintain convective neutrality ($\nabla_{\text{rad}} = \nabla_{\text{ad}}$) located above the convective core, as depicted in the upper panel of Figure~\ref{fig2}. The balance of He, C, and O in the semiconvective area is critical, as it dictates the necessary decrease in the radiative gradient within the region. The CPM scheme appears to be a promising substitute for the offered two numerical methods, which suffer from problems such as the appearance of the convective shell and having $\nabla_{\text{rad}} > \nabla_{\text{ad}}$ on the convective side of the convective core boundary (see the black plot in the upper panel of Figure~\ref{fig2}). As a result, we have decided to apply the CPM scheme in all models used in our research.

A thorough understanding of the precursor phases leading to the evolution and formation of sdB stars is essential for studying these stellar objects. This study employs the {\sc mesa} algorithm to model the evolution of single stars, as shown in Figure 1 of \citet{Ghasemi-2017}. The star evolves from the pre-main-sequence phase to the tip of the red giant branch (TRGB). The core of the star contracts during the RGB evolutionary process, which reduces the mean separation between the constituent particles. Subsequently, as the core contracts and releases gravitational energy, an electron-degenerate He core is formed. 

To replicate the effects of binary interaction and envelope mass stripping that lead to the formation of an sdB star, the {\tt relax\_mass} option in {\sc mesa} is utilized. This approach involves the incorporation of a high mass-loss rate starting at the TRGB.  After the process of envelope mass stripping, the resulting model resembles an sdB star, typically with a mass of around 0.47$M_{\odot}$ and a very low envelope mass, such as 0.002$M_{\odot}$. In the subsequent phase, the onset of successive core He flashes occurs. The weak interaction of neutrinos with matter facilitates the effective cooling of the central regions within the degenerate He core. As the He flashes occur, nearly 5 percent of the He fuses into C and O. The energy liberated by these nuclear processes elevates the core temperature, consequently augmenting the de Broglie wavelength of the electrons. This action removes the electron degeneracy from the He core. The inward progression of the He flashes continues over approximately 2 million years \citep{Bildsten-2012, Ghasemi-2017}. Ultimately the core contracts, stable helium burning through the triple-alpha reaction yields C during the EHB phase.

\begin{figure}[ht]
    \includegraphics[width=0.5\textwidth]{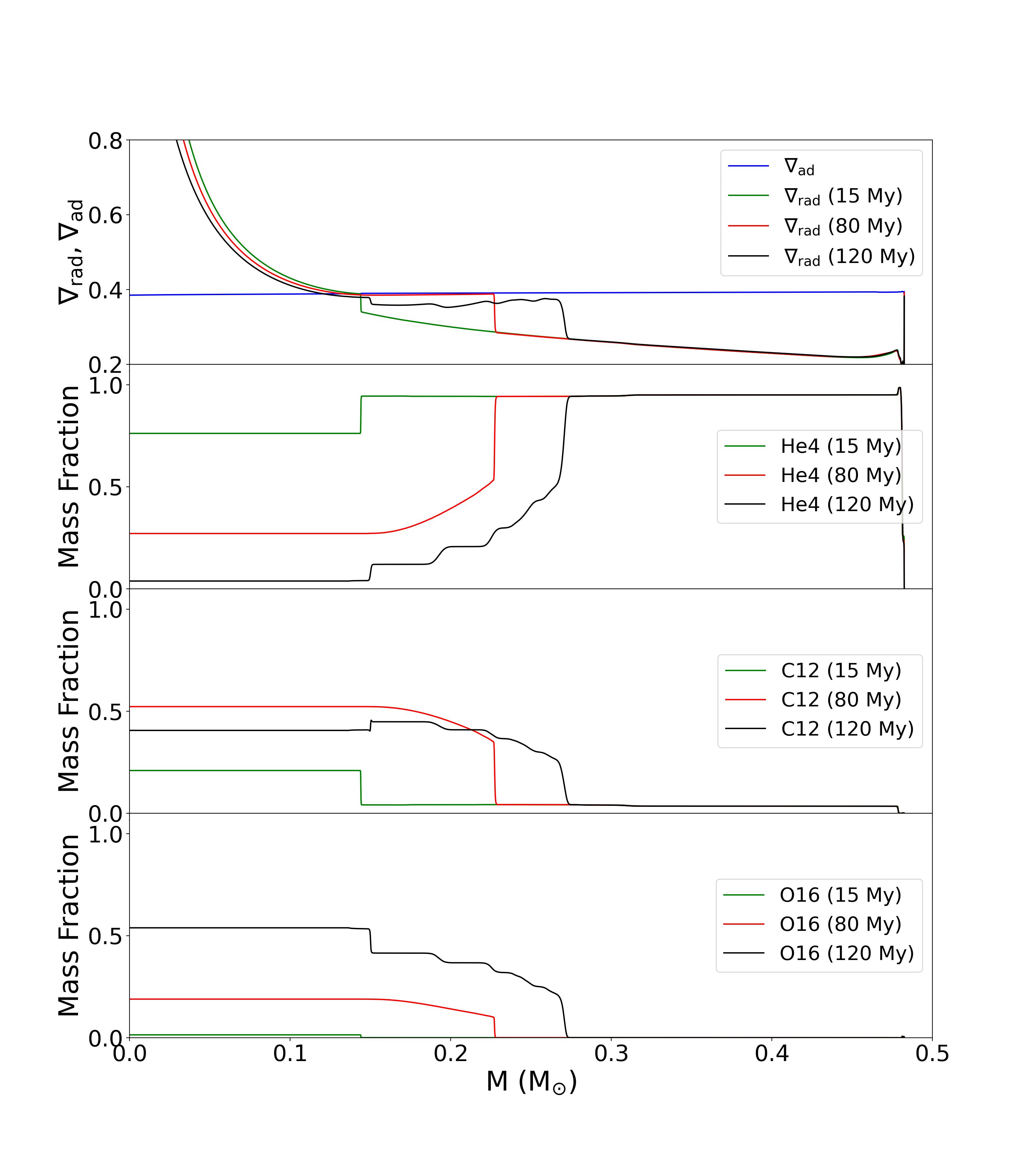}
    \caption{\textbf{Top Panel:} The adiabatic gradient is represented in blue, accompanied by radiative gradients for three distinct sdB models depicted in Figure~\ref{fig1c} all plotted as mass coordinates. These models are differentiated by their respective ages: 15 million years (My) shown in green, 75 My in red, and 120 My in black. The \textbf{bottom panels} show, from top to bottom, the He, C, and O mass fractions corresponding to each of the sdB models listed in the top panel.}
    \label{fig3}
\end{figure}

During the EHB phase, the convective core progresses in distinct stages, each with its own unique characteristics, instead of progressing uniformly. Overall, we can identify three distinct stages in its evolution. Calculations performed using another code, {\sc STELUM}, also reveal nearly the same three main stages during the helium-burning phase \citep{Giammichele-2022}. The initial stage spans from the zero-age extreme horizontal branch (ZAEHB) to the point where nearly 30\% of the core He is consumed. During this phase, the convective core expands consistently (that can be seen in stellar ages from 0 to roughly 20 million years in Figure~\ref{fig1c}), and there is no requirement for a semiconvective region. The neutrality condition ($\nabla_{\text{rad}} = \nabla_{\text{ad}}$) persists on the convective side of the core boundary, as depicted in the top panel of Figure~\ref{fig3}.

The second stage lasts until approximately 10\% of the He remains in the core (as seen in stellar ages from nearly 20 to roughly 110 million years in the bottom panel of Figure \ref{fig1}). To meet the neutrality condition ($\nabla_{\text{rad}} = \nabla_{\text{ad}}$), the quantities of He, C, and O must be adjusted appropriately at each point within the semiconvection region. This involves adjusting the chemical composition to maintain a temperature gradient that closely matches the adiabatic gradient, as illustrated in red in the top panels of Figure \ref{fig3}. This numerical approach succeeds well in the primary phase of central He burning when the amount of O produced is not yet substantial.  

The third stage is known as breathing pulses. These pulses arise from the efficient nuclear transformation of C into O. Since O has a higher opacity than C, this transformation leads to an increase in opacity, which in turn creates a higher radiative gradient, depicted in black in the top panel of Figure \ref{fig3}. As a result, the convective core expands largely, allowing for the mixing of large quantities of He into the burning zone. After this expansion, the core contracts over a short evolutionary timescale, leading to several evolutionary phases marked by smaller, growing convective cores that resemble the initial stage, free from the semiconvective region, as shown in Figure \ref{fig1c} at the end of core He-burning phase. The breathing pulses are recurring events significantly influenced by mixing processes near the convective core and by the numerical methods used to apply these processes at the convective boundary. If the boundary of the convective core reaches the outer layers of He at sufficiently low temperatures, pulses can occur before the nuclear fuel in the core is depleted. Otherwise, the He-burning phase ends.

\section{Deciphering the asteroseismic properties of sdB models}
\label{sec:3}

According to asymptotic analysis, high-order g modes should have periods that are equally spaced in chemically homogeneous non-rotating non-magnetic stars \citep{Tassoul-1980}. However, the measured period spacings of g modes are known to be non-uniform. These deviations can be employed to more thoroughly understand the properties of the stellar interior, particularly the interior chemical mixing characteristics \citep{Miglio-2008}.

In order to comprehend the structure of compact evolved stars, the period spacing between p and g-modes has been studied. The asymptotic theory analysis of white dwarfs with an outer convection zone and a discontinuity in their chemical composition was developed by \citet{Brassard-1992a, Brassard-1992b}. The asymptotic theory for the case of sdB stars with a convective core and a radiative envelope was introduced by \citet{Charpinet-2000, Charpinet-2002a, Charpinet-2002b}.

\begin{figure}[ht]
    \centering
    \subfigure[]{
        \includegraphics[width=0.5\textwidth]{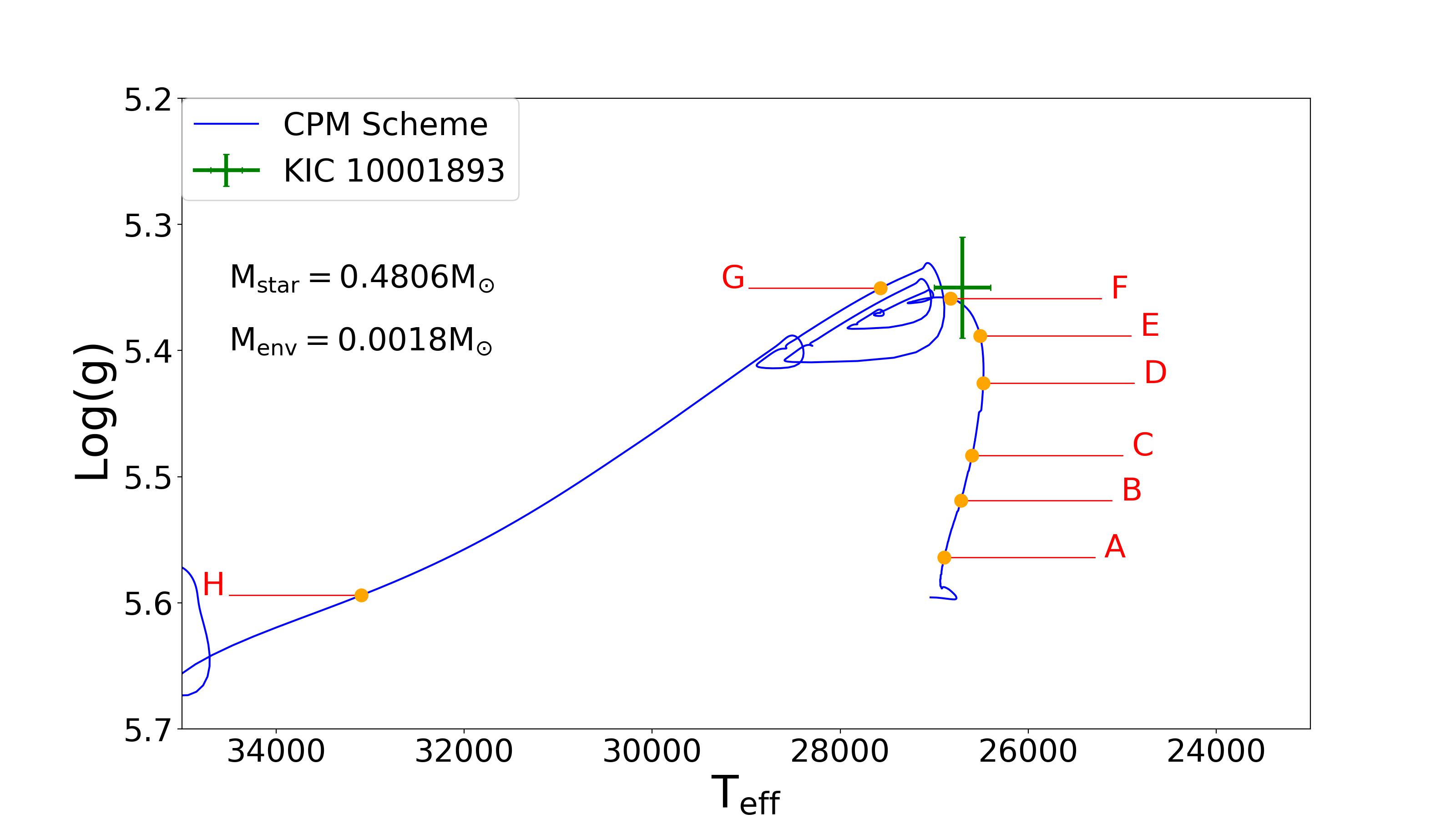}
        \label{fig4a}
    }
    \hfill
    \subfigure[]{
        \includegraphics[width=0.5\textwidth]{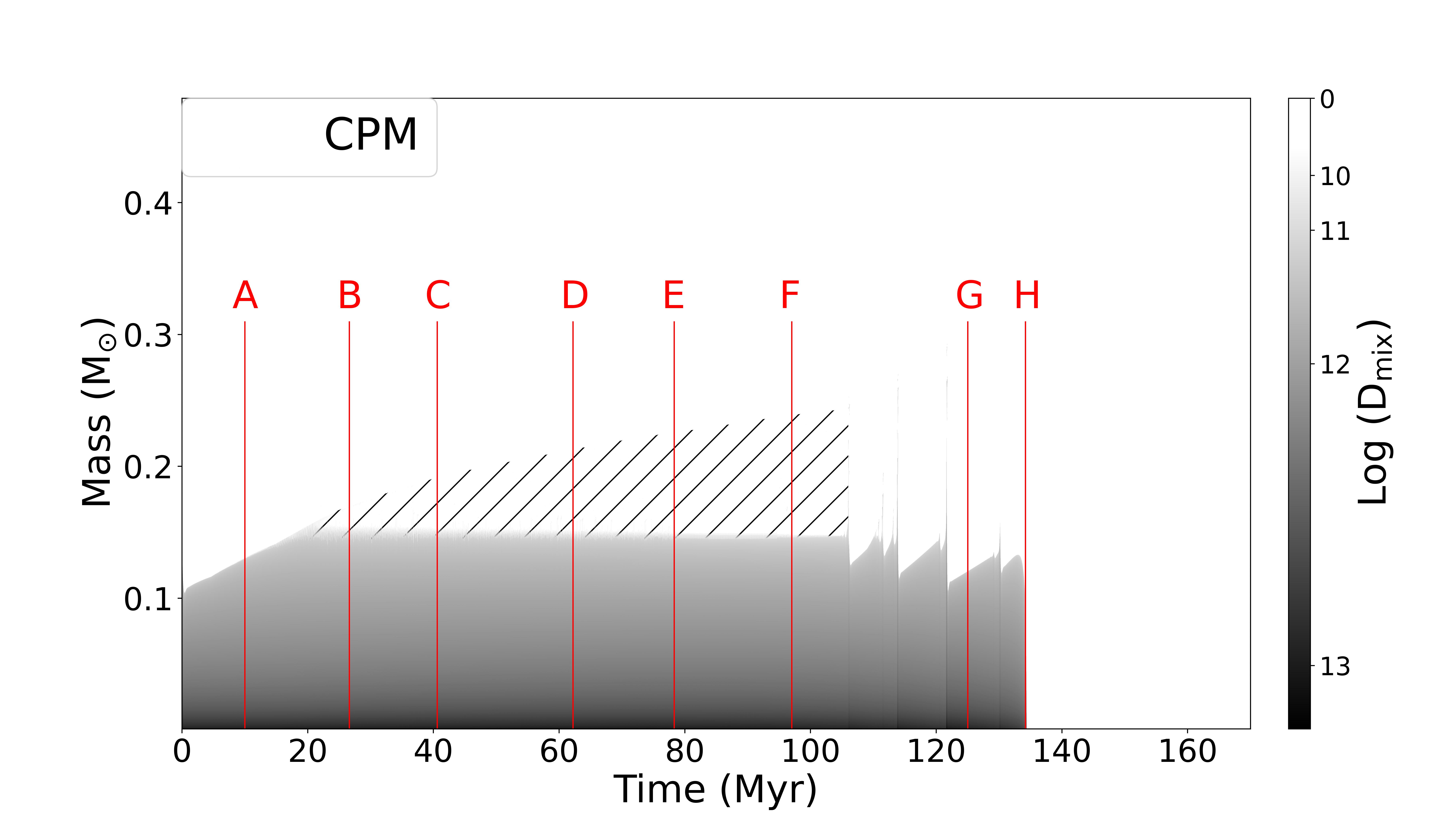}
        \label{fig4b}
    }
    \caption{\textbf{Panel (a)} Kiel diagram (stellar gravity ($\log g$) against effective temperature ($T_{\text{eff}}$) for the CPM scheme. The position of KIC 10001893 is illustrated in green with error bars. Eight models with different ages (A=10 My, B=27, C=41, D=62, E=78, F=97, G=125, H=134 million years) are indicated. \textbf{Panel (b)} is similar to panel (C) in Figure~\ref{fig1}. Hatched areas indicate semiconvection in the transitional zone between the convective (gray) and radiative (white) regions. All eight models from the top panel are also indicated in this panel.}
    \label{fig4}
\end{figure}

In Figure \ref{fig4a}, the Kiel diagram displays the relationship between stellar gravities ($\log g$) and effective temperatures ($T_{\text{eff}}$) for the CPM scheme. The spectroscopic properties of KIC 10001893 are depicted in green, accompanied by error bars. Eight models labeled A to H with varying ages (10, 27, 41, 62, 78, 97, 125, and 134 million years) represent all three evolutionary stages discussed earlier. Model A represents the initial uniform growing convective core phase, while models B to F mark the second evolutionary phase with a semiconvection region. Model G corresponds to the breathing pulses phase, and point H denotes the end of the He-burning phase. These models are also featured in Figure \ref{fig4b}, which illustrates the schematic representation of sdB star interior structures during the He-burning stage using the CPM scheme.

Different mixing methods yield unique configurations of the chemical gradient around the convective core, impacting the period spacing pattern. Therefore, the observed period spacing patterns provide the potential to improve our understanding of the mixing dynamics near the convective core and the effectiveness of the mixing mechanisms.

The boundary layer between the C/O core and the He shell has a significant impact on the period spacing patterns. Moreover, the semiconvection zone depicted in light gray in Figure \ref{fig4b} can generate deep mode trapping patterns. A sharp composition gradient marks the boundary between the semiconvection zone and the outer radiative layer. The semiconvection region occupies the space between this boundary and the lower boundary of the convective core.

\begin{figure}[ht]
    \includegraphics[width=0.5\textwidth]{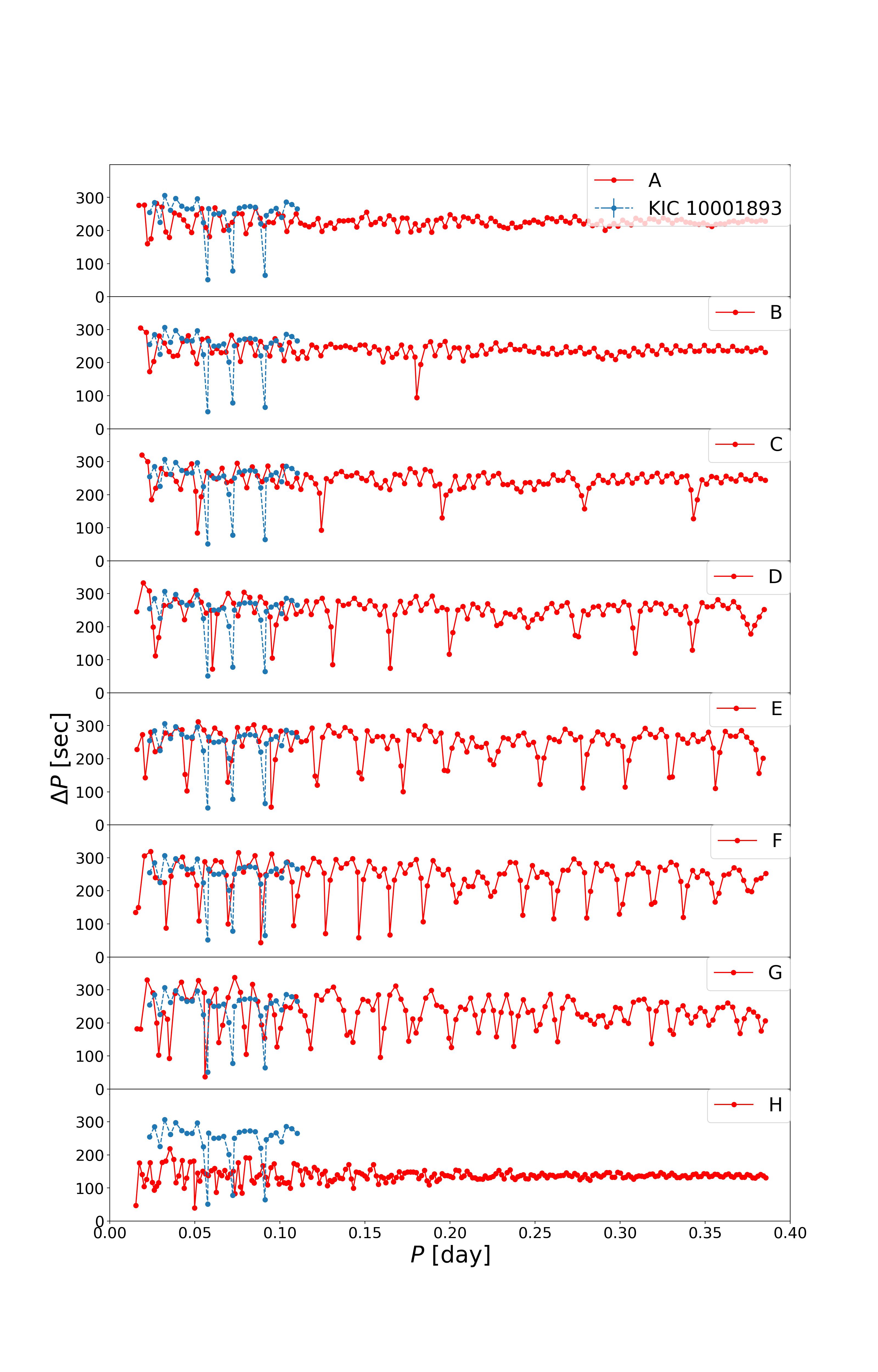}
    \caption{The measured period spacing of KIC 10001893 is represented in blue, while the period spacings of the eight models depicted in Figure~\ref{fig4} are shown in red. The order of the figures corresponds to their age, with the youngest at the top and the oldest at the bottom.}
    \label{fig5}
\end{figure}

Figure \ref{fig5} illustrates the period spacing of KIC\,10001893 in blue, contrasted with the red period spacings of the eight models featured in Figure \ref{fig4}. The sequence of panels aligns with their respective ages, with the youngest positioned at the top and the oldest at the bottom. The semiconvection generates intricate deep-mode trapping patterns. As the evolutionary path progresses, the distance between the sharp composition gradients in the outer semiconvection area and the lower convective core increases, leading to an increase in the number of trapped modes. Consequently, at model A, where the semiconvection area is absent, deep trapped modes do not exist, and the period spacing solely relies on the He/H transition layer and the sharp composition gradient at the top of the convective core boundary. As the size of the semiconvection region increases from models B through F, the number of deeply trapped modes also rises. Furthermore, the spacing between these trapped modes remains relatively constant over this range. However, this pattern changes at models G and H, where the distances between the deeply trapped modes are no longer uniform.

\begin{figure}[ht]
    \centering
    \subfigure[]{
        \includegraphics[width=0.5\textwidth]{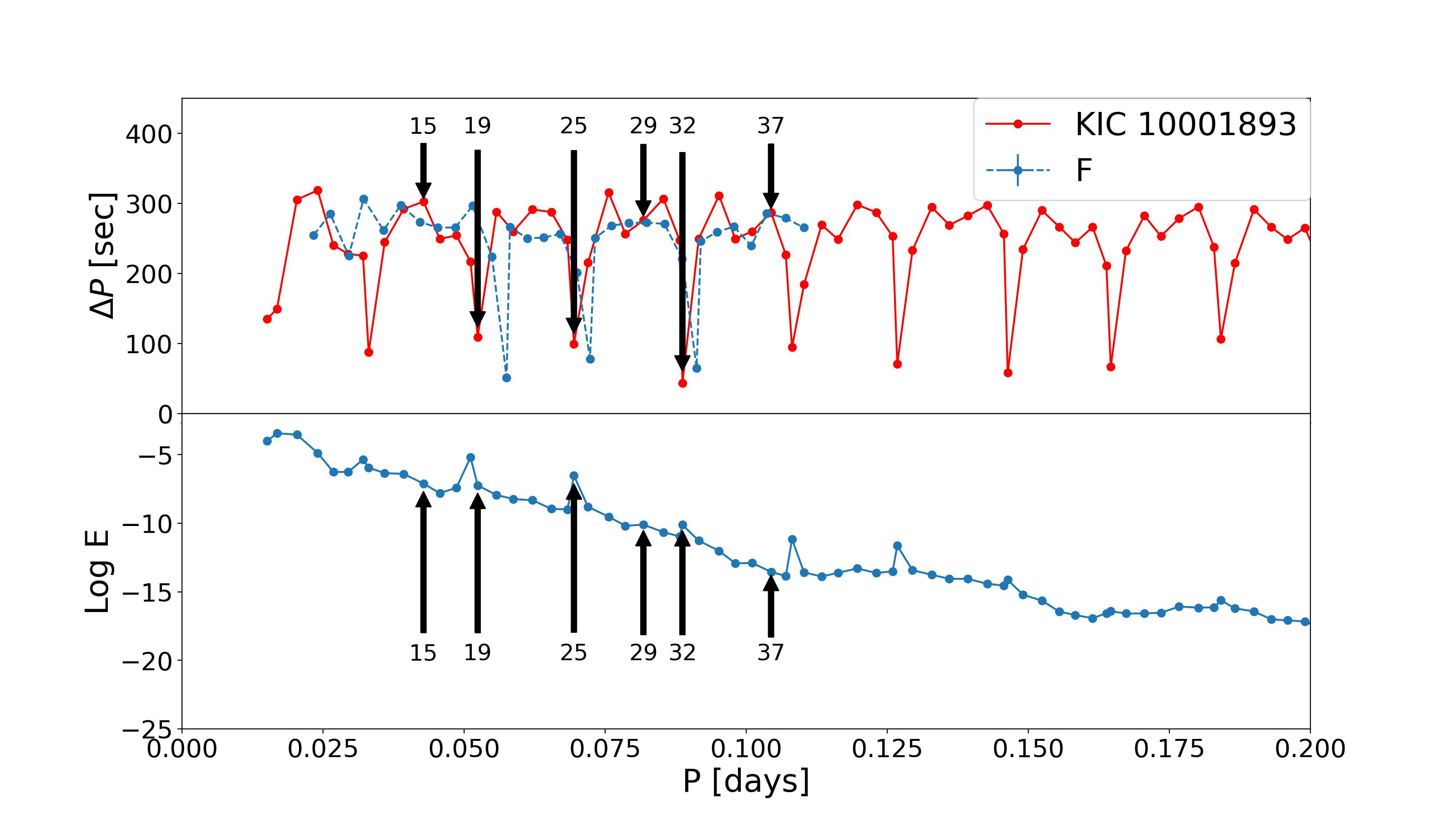}
        \label{fig6a}
    }
    \hfill
    \subfigure[]{
        \includegraphics[width=0.5\textwidth]{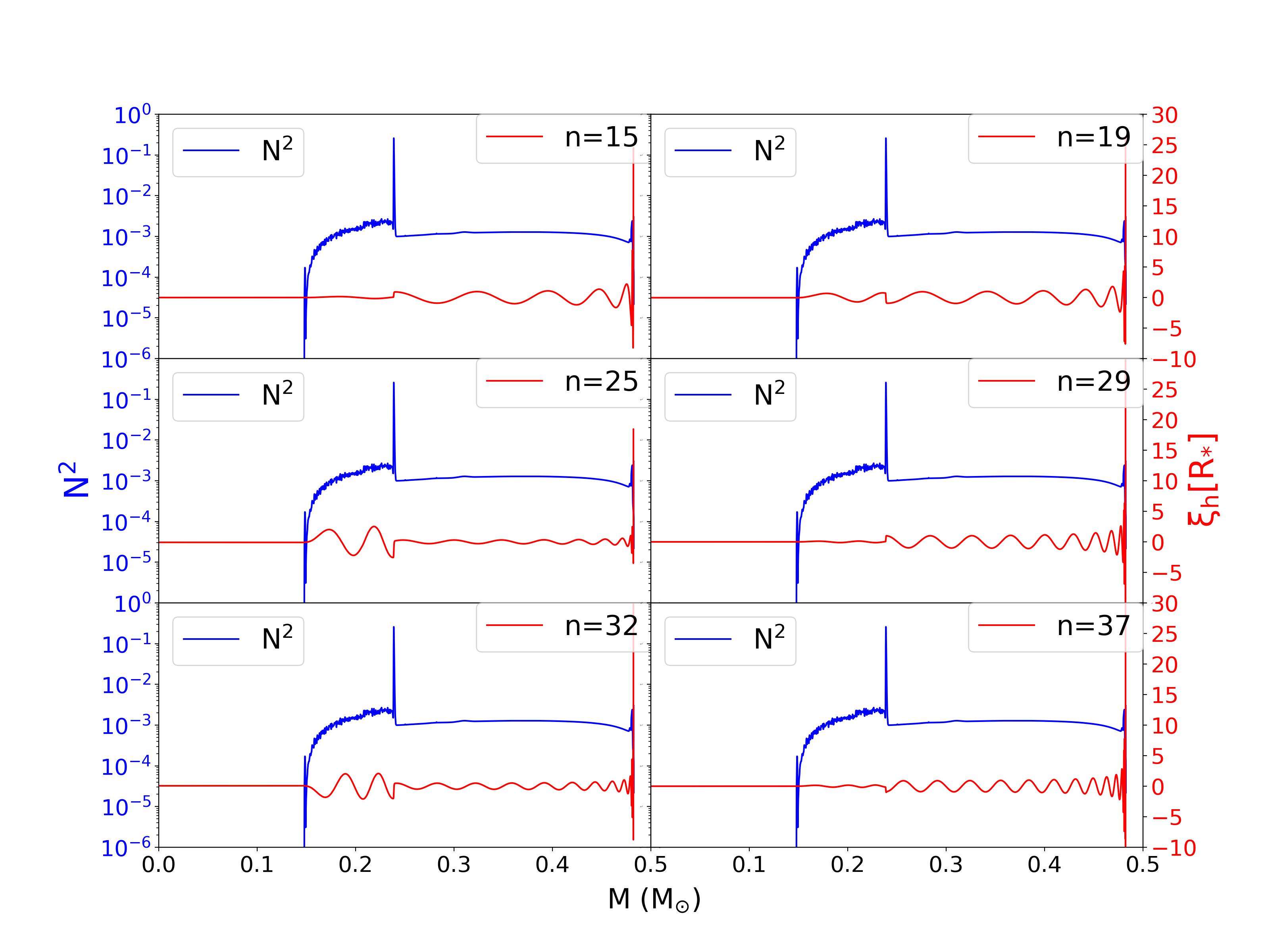}
        \label{fig6b}
    }
    \caption{\textbf{Panel (a)} (Top): Period spacing pattern of g-modes with degree $\ell = 1$ and consecutive radial order \( n \) for a stellar model employing the CPM scheme, corresponding to point F in Figure ~\ref{fig4}. Radial orders for six modes are indicated by black-filled arrows.
    \textbf{Panel (a)} (Bottom): Normalized mode inertia plotted against the periods of the dipole (\( \ell = 1 \)) g-modes.
    \textbf{Panel (b)}: Horizontal displacement eigenfunctions $(\xi_h)$ of the dipole g modes plotted as a function of mass coordinate for various radial orders, as referenced in the above panels. Red solid lines represent the eigenfunctions, while blue solid lines depict the Brunt-Väisälä frequency within the stellar interior.}
    \label{fig6}
\end{figure}

Figure~\ref{fig6a} (Top) illustrates the periodic spacing pattern of g-modes with a degree of $\ell = 1$ and consecutive radial order \( n\) for a stellar model employing the CPM scheme, corresponding to the model F in Figures ~\ref{fig4} and ~\ref{fig5}. The radial orders of six modes are denoted by black-filled arrows. Changes in the chemical gradients alter the density profile and, consequently, the Brunt-Väisälä frequency. Within the sdB models, transitions from H to He in the outer envelope and from He to C/O near the convective boundary create distinct features in the Brunt–Väisälä frequency, as indicated by the blue solid line in Figure ~\ref{fig6b}. The analysis reveals that the horizontal component of the eigendisplacements \(\xi_h \), depicted by the red solid line in Figure ~\ref{fig6b} for modes with radial orders \(n=15, 29, 37\), are confined in the vicinity of the Brunt–Väisälä frequency peak near the core. Certain displacement eigenfunctions, with nodes situated very close to the semiconvective zones, exhibit significant amplitudes inside the semiconvection regions, shown by the red solid line in Figure~\ref{fig6b} for modes with radial orders \(n=19, 25, 32\). 

The normalised inertia of these modes represents the average kinetic energy associated with displacement eigenfunctions across time. Figure~\ref{fig6a} (Bottom) shows the normalised mode inertia plotted against the period of the dipole $(\ell = 1)$ g-modes. In the CPM scheme, a few modes with large amplitudes become trapped within the semiconvective area. Consequently, these deep trapped modes display significant mode inertias, as indicated by the filled black arrows in Figure~\ref{fig6a} (Bottom), for modes with radial orders \(n=19, 25, 32\).

One of the lower panels in Figure~\ref{fig5} illustrates the period spacing of model G, which is associated with the breathing pulses. Seismic measurements of the central oxygen abundance in white dwarfs point to the occurrence of breathing pulses in the cores of their progenitors during the helium-burning phase \citep{Giammichele-2022}. This highlights the importance of breathing pulses in stellar models. In this phase, as C undergoes efficient conversion into O within the core, the opacity increases, leading to a rise in the radiative gradient. This, in turn, triggers a sudden expansion of the convective core, followed by the emergence of a Brunt–Väisälä frequency peak in the outer radiative region. Subsequently, there is a contraction, resulting in the appearance of Brunt–Väisälä frequency peaks in the inner radiative region, eventually leading to the formation of a small growing convective core. The presence of multiple Brunt–Väisälä frequency peaks above the expanding convective core can give rise to irregular patterns in trapped modes, where modes with nodes close to these peaks become trapped. Due to the varied distances between these peaks, the average interval between the trapped modes and the trapping depth of these modes may differ, causing irregularities in the pattern of trapped modes. At model H in Figure~\ref{fig5}, which corresponds to the end of He burning, the period spacing decreases as the convective core size shrinks.

\begin{figure}[ht]
    \includegraphics[width=0.5\textwidth]{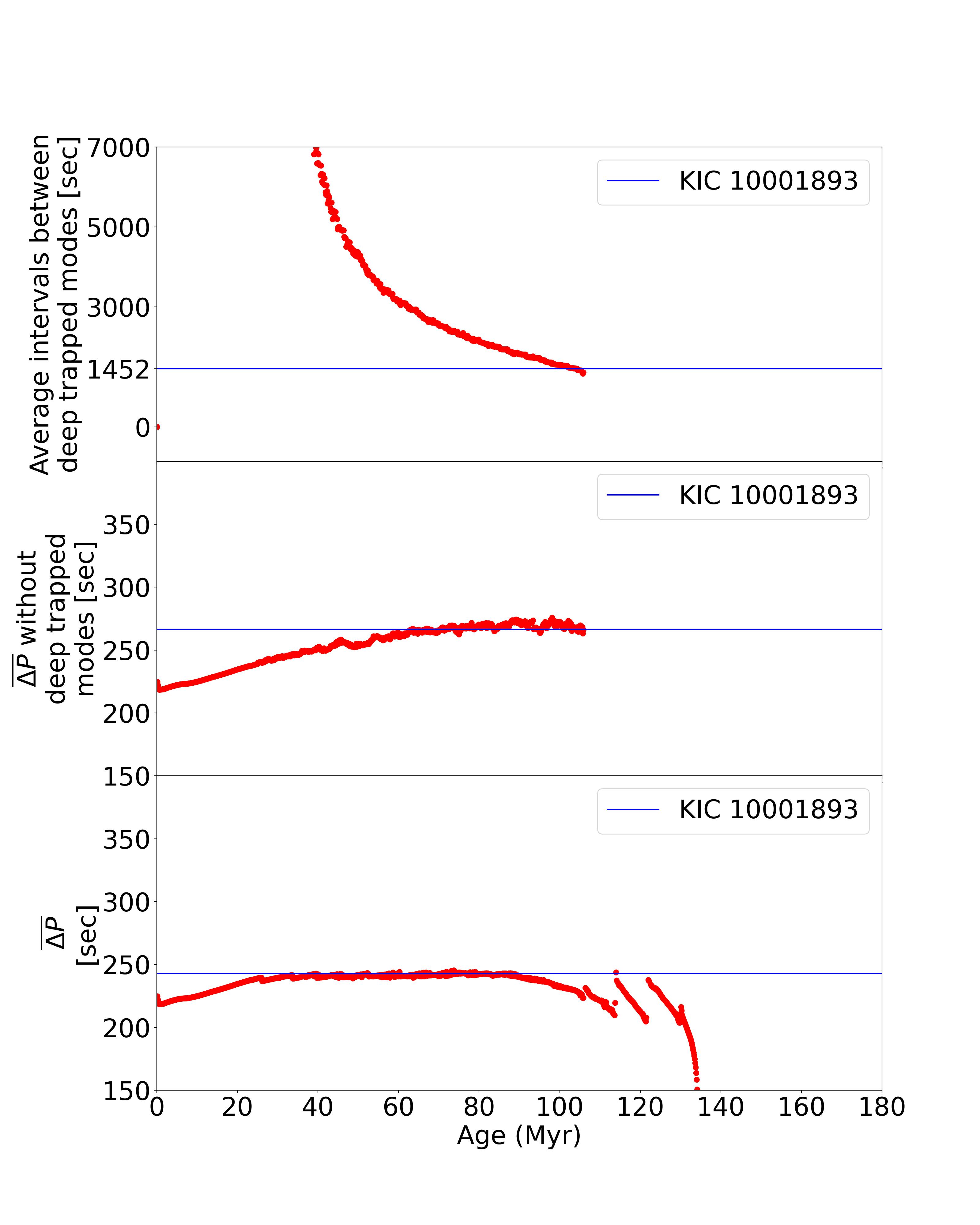}
    \caption{\textbf{Top Panel}: The parameter indicates the average interval between the deep trapped modes during the second stage of evolving the sdB models for the CPM scheme, as shown in Figure~\ref{fig4b} by red dots, and the observed mean distances between three deep trapped modes of KIC 10001893, depicted with a blue line.
     \textbf{Middle Panel}: The mean period spacing without considering the deep trapped modes in the first and second evolutionary stages of the sdB models, illustrated by red dots. The blue line represents the observed mean period spacing, excluding three deep trapped modes and three semi-trapped modes of KIC 10001893.
     \textbf{Bottom Panel}: The mean period spacing of all modes in all the evolutionary stages of He burning of the sdB models, as depicted by red dots. The blue line displays the observed mean period spacing of all modes of KIC 10001893.}
    \label{fig7}
\end{figure}

The red dots in the top panel of Figure~\ref{fig7} illustrate a new parameter that signifies the average interval between deep trapped modes during the second stage of the sdB evolution, as elaborated in Section~\ref{sec:2}. The observed mean distances between three deep trapped modes of KIC 10001893 are denoted by a blue line. Moving to the middle panel of Figure~\ref{fig7}, the red dots display the mean period spacings excluding deep trapped modes in both the first and second evolutionary stages of the sdB evolution, as elaborated in Section~\ref{sec:2}. Meanwhile, the observed mean period spacing of KIC\,10001893, excluding three deep trapped modes and three semi-trapped modes, is represented by a blue line. Lastly, the bottom panel of Figure~\ref{fig7} showcases the mean period spacing throughout all evolutionary stages of He burning of the sdB, as illustrated by red dots. The observed mean period spacing of KIC 10001893 is displayed by a blue line.

As shown in the lower panel of Figure~\ref{fig7}, it becomes apparent that using the mean period spacing of all modes as a parameter to define the core He-burning phase during the second stage of sdB evolution is impractical. Throughout the majority of this phase, the mean period spacing of all modes remains relatively constant. Even when considering the mean period spacing for modes not trapped in the semiconvection region, without accounting for the deep trapped modes and semi-trapped modes in this region, discerning a clear pattern for the last part of the second stage of He burning becomes challenging. This is primarily due to the inconsistency in the stability of the convective core boundary, resulting in fluctuations in the mean period spacing, as depicted in the middle panel of Figure~\ref{fig7}. Furthermore, As shown in the lower panel of Figure~\ref{fig7}, it is clear that using the mean period spacing during the first stage of sdB evolution remains reliable.

In our sdB models, we identify deep trapped modes within the semiconvective region by their horizontal displacement, which are required to exceed 0.5 $R_{*}$ within that region. As shown in Figure ~\ref{fig6b}, the modes with radial orders 19, 25, and 32 all have horizontal displacements above this threshold within the semiconvective zones. As depicted in the top panel of Figure~\ref{fig7}, utilizing the average interval parameter between deep trapped modes proves to be an effective means of distinguishing the core He-burning phase. Throughout the second stage of He burning, the average interval parameter exhibits a consistent decrease as the age increases, displaying a steady decline without fluctuations. Consequently, the average interval parameter emerges as a reliable metric for characterizing the He-burning phase within our sdB evolutionary models. Notably, we refrain from specifying the average interval parameter for the third evolutionary stage, the breathing pulses, due to the almost negligible presence of the semiconvection region and the irregular pattern of the trapped modes during this phase.

\section{Time-dependent Convective Penetration}
\label{sec:4}
\citet{Johnston-2024} described how the dissipation balanced convective penetration procedure, initially developed in 3D numerical simulations by \citet{Anders-2022}, was implemented into the {\sc mesa} 1D stellar evolution model. \citet{Jermyn-2022} employed a similar approach to estimate the penetration zone (PZ) in the post-processing step while ignoring the effects of the PZ on the evolution of stars.

Multidimensional hydrodynamic simulations consistently reveal the emergence of a convective penetration zone, where the temperature gradient shifts smoothly from $\nabla_{\text{rad}}$ to $\nabla_{\text{ad}}$. In this section, we utilize this methodology alongside the CPM scheme to demonstrate evolutionary and asteroseismic outcomes when incorporating mixing beyond the convective core without introducing additional free parameters to our models. This algorithm is founded on the concept that convective parcels extend beyond boundaries due to velocity and inertia, leading to convective penetration. The algorithm calculates the extent of this penetration zone above the convective boundary by balancing buoyant work against dissipative forces, assuming full mixing with an adiabatic temperature gradient. This time-dependent convective penetration algorithm calculates the extent of the penetration zone at each time step of the evolution using the properties of the star, unlike previous methods, which remove free parameters and determine the penetration zone according to the properties of the model as a post-processing step \citep{Jermyn-2022}.

The insights from 3D hydro simulations are derived, as described by \citet{Anders-2022}, to characterize the dissipation profile within the PZ and the dissipation-to-buoyant work ratio in the convective zone. These findings are then included in the 1D {\sc mesa} code. Furthermore, the temperature gradient within the convective zone and PZ is adjusted to replicate the adiabatic gradient and smoothly transition to the radiative gradient via the Péclet number \citep{Michielsen-2021}. The Péclet number is a dimensionless measure that compares the thermal diffusion timescale to the convective turnover time to describe the relative importance of convective and radiative heat transport \citep{Jermyn-2022}.

\begin{figure}[ht]
    \centering
    \subfigure[]{
        \includegraphics[width=0.5\textwidth]{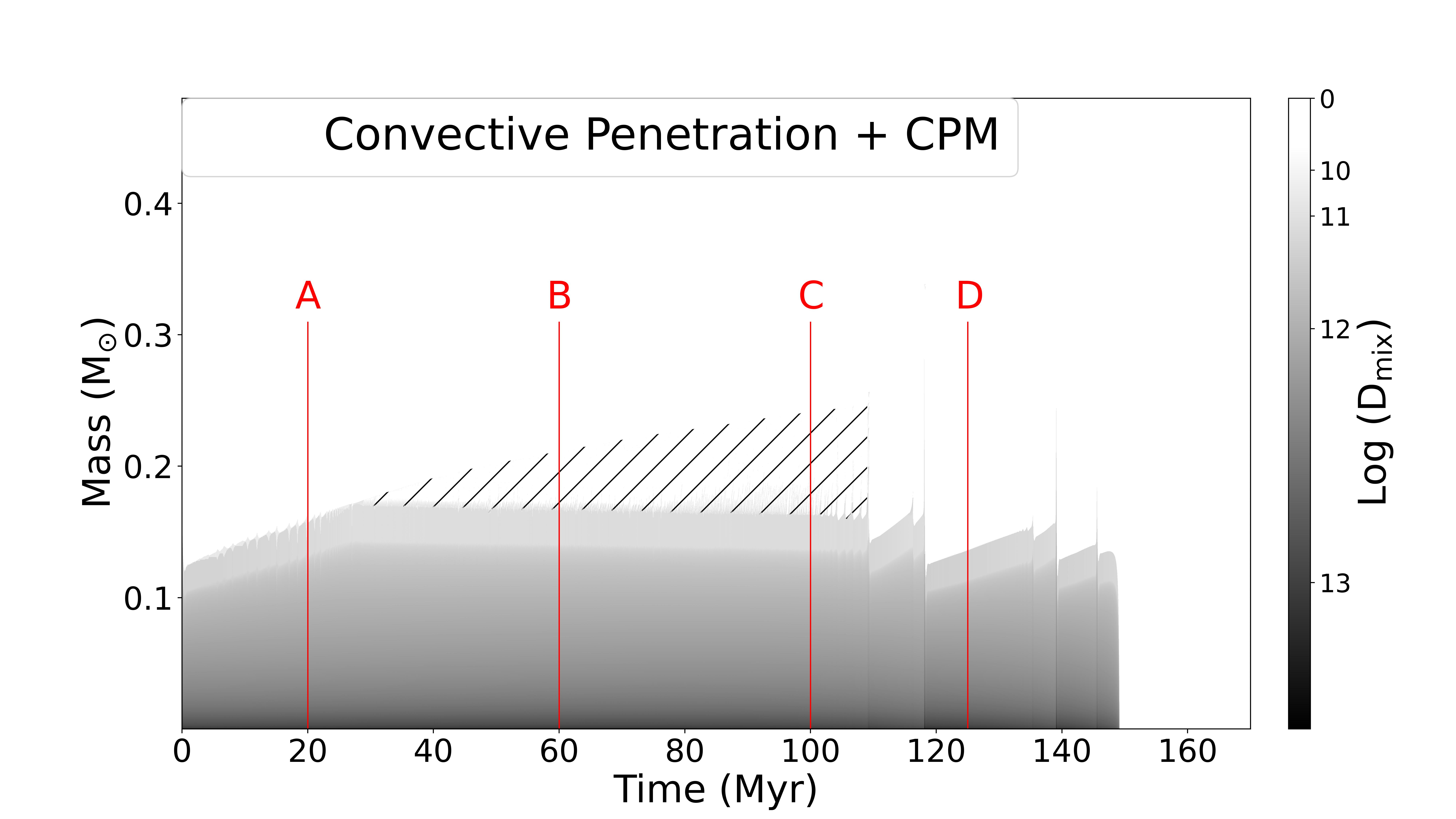}
        \label{fig8a}
    }
    \hfill
    \subfigure[]{
        \includegraphics[width=0.5\textwidth]{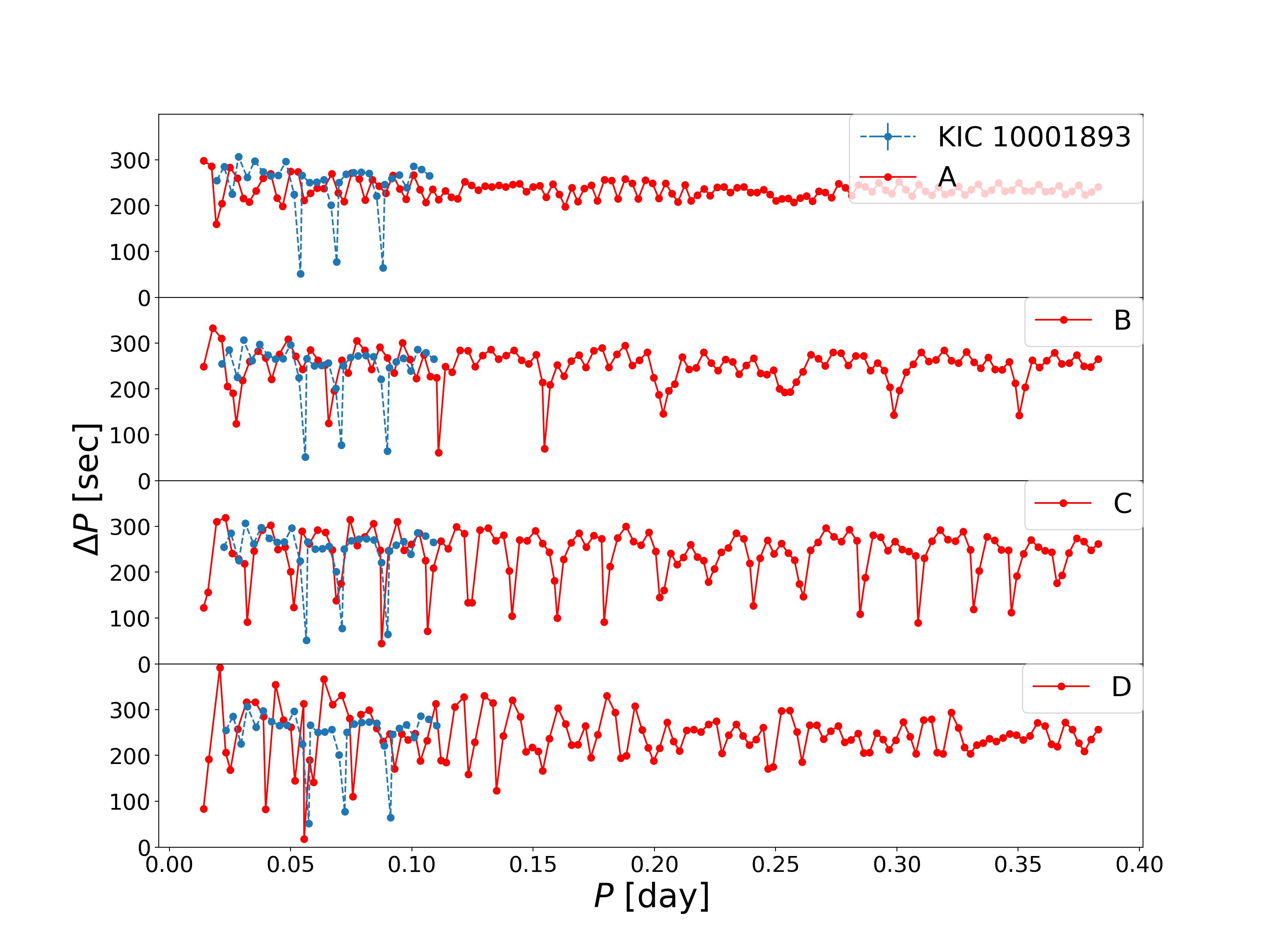}
        \label{fig8b}
    }
    \caption{\textbf{Panel (a)} depicts a Schematic representation of the sdB star interior structures for the logarithm of the diffusion coefficient of mixing during the core He-burning stage, using convective penetration combined with the CPM scheme. Different levels of mixing are indicated by the gray-scale bar on the right-hand y-axis. The hatched area represents semiconvection in the transitional zone between the convective (gray) and radiative (white) regions. The figure shows four points with varied ages (A=20, B=60, C=100, and D=127 million years). \textbf{panel (b)} displays the period spacing of KIC 10001893 in blue, while the period spacings of the four models depicted in panel (a) are shown in red. The order of the figures corresponds to their ages, with the youngest at the top and the oldest at the bottom.}
    \label{fig8}
\end{figure}

The Figure~\ref{fig8a} provides a schematic representation of the internal structures of the sdB star during the He-burning phase, highlighting the phenomenon of convective penetration. This region is shown in a lighter gray shade compared to the convective core region. The hatched area represents semiconvection in the transitional zone between the convective (gray) and radiative (white) regions. This figure features four distinct models representing different ages (A=19, B=68, C=100, and D=118 million years). In Figure~\ref{fig8b}, the blue lines indicate the period spacing of KIC 10001893, while the red lines represent the period spacings of the four models showcased in Figure~\ref{fig8a}. These panels in Figure~\ref{fig8b} are arranged in descending order of age, with the youngest depicted at the top and the oldest at the bottom.

Figure~\ref{fig8a} illustrates the convective penetration situated between the convective core and the radiative and semiconvective regions. Alternatively, the convective penetration may be substituted with overshooting, or the overshooting can be introduced above the convective penetration (see \citet{Michielsen-2021}), yielding similar outcomes. In Figure~\ref{fig8b}, the period spacings of the models exhibit a consistent pattern reminiscent of those observed in the CPM scheme, albeit without the presence of convective penetration as shown in Figures~\ref{fig4} and~\ref{fig5}. Most modes, except for the deep trapped ones, reflect off the outer boundary of the semiconvective region. Only the deep trapped modes penetrate this region, therefore the mean period spacing for non-trapped modes depends on the size of this boundary, which is the same for both model C in Figure~\ref{fig8a} and model F in Figure~\ref{fig4b}. On the other hand, the size of the semiconvective region reflects the depth of the deep trapped modes. As our models age, the semiconvective region grows. Since the depth of the deep trapped modes is nearly identical in the two models, model C in Figure~\ref{fig8a} which has a larger convective core due to its convective penetration region, occurs later, around 100 million years, compared to model F in Figure~\ref{fig4b}, which occurs at approximately 97 million years. 

At the end of the He-burning phase, due to the different number and sizes of breathing pulses, the fully enriched C/O core and the abundance gradient region below the fully He-rich layers differ in size and compositional structure between the models in Figure~\ref{fig4b} and those in Figure~\ref{fig8a}.

\section{Discussion and conclusion}
\label{sec:5}
Our research focused on the evolution of the convective core in sdB stars, revealing a non-uniform progression marked by three distinct stages in the CPM scheme \citep{Paxton_2019, Ostrowski-2021}. The CPM scheme uses a numerical approach to model the formation of the adiabatic semiconvection zone during the core He-burning phase \citep{castellani1985,Mowlavi_1994}. Static asteroseismic models predict convective core masses significantly larger than those suggested by the CPM scheme {\citep{VanGrootel-2010a, VanGrootel-2010b,Charpinet-2011, Charpinet-2019}. In these models, the He/H transition layer and the deep core boundary layers play a key role in mode trapping as discussed extensively by \citep{Guyot-2025}, to which we refer for details. In the models examined in this study, variations in chemical composition caused by the CPM scheme within the transition zones near the cores of sdB stars lead to certain modes becoming strongly trapped, displaying similar depths and intervals. This phenomenon is supported by Kepler space-telescope observations of particular sdB stars \citep{ostensen2014, Uzundag2021}.

\citet{Gabriel_2014} suggested that in the context of the sign-change numerical approach, the persistent dominance of $\nabla_{\text{rad}} > \nabla_{\text{ad}}$ on the convective side of the core boundary underscores the need for additional mixing mechanisms such as adiabatic semiconvection. The sign-change can anticipate the emergence of multiple peaks in the Brunt–Väisälä frequency above the expanding convective core during specific phases of the second evolutionary stage. Such occurrences could lead to irregularities in the intervals and depths of the deep trapped modes \citep{Ghasemi-2017}. In the second evolutionary stage, the premixing technique predicts a single Brunt–Väisälä peak at the outer boundary of the semiconvection region to trap modes. Consequently, the CPM scheme yields nearly identical estimates for the average interval between deep trapped modes and the depth of all these modes.  According to \citet{Uzundag2017}, observations support the presence of three deep trapped modes, as confirmed by the mode period spacing with $\ell = 1$, and two of these deep trapped modes reconfirmed with $\ell = 2$, exhibiting identical depths, consistent with the CPM scheme predictions.

\cite{Blouin-2024} conducted a 3D simulation study of the He-burning phase in a 3 $M_{\odot}$ red clump star, exploring CPM and PM numerical approaches at the convective core boundary. However, the semiconvection region between the convective core and radiative envelope exhibits insufficient mixing due to internal gravity waves, contrary to the claim by \cite{Herwig-2023}. Nonetheless, their simulation suggests that convective intrusion and overshooting could potentially eliminate this region within just 10 to 50 convective turnover timescales. Extrapolating entrainment over evolutionary timescales might further erase this semi-convection region. In scenarios involving the penetration zone and the application of the CPM, employing higher mesh densities can introduce fluctuations during the second stage of helium burning a phase where our study anticipates smooth growth in semiconvective regions. After these fluctuations, the semiconvective region reforms, suggesting that the simulation in \cite{Blouin-2024} might be based on one of these minor variations. Additionally, according to \cite{Blouin-2024}, an incomplete representation could lead to an overestimation of the rate at which semiconvection is eroded. Most importantly, based on this study, the PM numerical approach results are not consistent with the g-mode asteroseismology of sdB stars and cannot explain the regions that cause deep trapped g-modes. Multidimensional hydrodynamics simulations consistently indicate the formation of a convective penetration zone, where the temperature gradient smoothly transitions from $\nabla_{\text{rad}}$ to $\nabla_{\text{ad}}$. This behaviour aligns with the modelling we demonstrated and discussed in Section~\ref{sec:4}.

Variations in the initial masses of sdB stars during their main sequence and red giant phases, combined with mass loss due to binary system interactions, can result in a wide range of total and envelope masses \citep{Hu-2007, Hu-2008}. This, in turn, influences the structure of the transition layer between He and H near the envelope, leading to variations in the Brunt-Väisälä peak. As the initial mass escalates from 1 $M_{\odot}$ to 2.5 $M_{\odot}$, the transition layer between He and H becomes smoother and broader, resulting in a lower and broader peak in the Brunt-Väisälä frequency. Consequently, there are fewer and less noticeable ordinary trapped modes in the period spacing pattern. However, the pattern of deep trapped modes remains the same.

Our future research will focus on constructing comprehensive grids of evolutionary models for sdB stars, which is crucial for determining the parameters of observed pulsating sdB stars. By testing various envelope masses and structures of the He-H transition layer, along with the CPM scheme above the convective core as a reliable model, we can improve the accuracy of pulsation predictions. The growing number of observed pulsating sdB stars, particularly with continuous high-precision space photometry, holds promise for identifying an increased proportion of sdB stars exhibiting trapped modes. This advancement can empower us to harness the pulsations of sdB stars, refining the determination of parameters for observed pulsating sdB stars, particularly when encountering trapped mode patterns. In addition, observing more sdB stars in clusters can enhance the accuracy of determining their parameters by incorporating the CPM scheme. Since pulsating sdB stars within a single cluster share the same distance, and composition, they offer additional constraints for parameter determination through asteroseismic grid modelling. The ages inferred from the ZAMS are approximately the same for all sdB stars within a cluster. However, it is important to note that the ages differ when determined from the onset of core helium burning as the stars within a cluster do not evolve into sdB stars at one particular time. Rather, this onset is determined by the birth mass of the star. Moreover, the ratio of sdB stars to sdO stars in clusters aids in estimating the lifetime of sdB stars during the core helium-burning phase relative to the lifetime of sdO stars in the shell helium-burning phase. This estimation heavily relies on the overall duration of the breathing pulses and the accuracy of the CPM scheme.

\begin{acknowledgements}

M. U. gratefully acknowledges funding from the Research Foundation Flanders (FWO) by means of a junior postdoctoral fellowship (grant agreement No. 1247624N). C.J.~acknowledges funding from the Royal Society through the Newton International Fellowship funding scheme (project No.NIF$\backslash$R1$\backslash$242552). CA received funding from the European Research Council (ERC) under the Horizon Europe programme (Synergy Grant agreement N$^\circ$101071505: 4D-STAR).  While partially funded by the European Union, views and opinions expressed are however those of the author(s) only and do not necessarily reflect those of the European Union or the European Research Council. Neither the European Union nor the granting authority can be held responsible for them.

\end{acknowledgements}

\bibliographystyle{aa}
\bibliography{myrefs}

\begin{appendix}
\section{Computational Models of sdB Stars}
\label{app:a}
We employ version r23.05.1 of the {\sc mesa} code (Modules for Experiments in Stellar Astrophysics; \citet{Paxton_2011, Paxton_2013, Paxton_2015, Paxton_2018, Paxton_2019, Jermyn_2023} to compute the evolutionary models of sdB stars. We use {\sc gyre} version 7.2.1 to calculate the adiabatic eigenfrequencies and eigenfunctions for the normal oscillation modes of the {\sc mesa} models \citep{Townsend-2013, Townsend-2018}. The following input physics are used to evolve an sdB model from the pre-main-sequence phase to the end of core He burning in the EHB phase. The initial stellar mass used in our modelling of sdB stars, during the pre-main-sequence and main-sequence phases, is 1.5 $M_{\odot}$. 

The {\sc mesa} EOS incorporates a combination of the OPAL \citep{Rogers2002}, SCVH \citep{Saumon1995}, FreeEOS \citep{Irwin2004}, HELM \citep{Timmes2000}, PC \citep{Potekhin2010}, and Skye \citep{Jermyn2021} equations of state (EOSes). Radiative opacities are primarily derived from OPAL \citep{Iglesias1993, Iglesias1996}, supplemented by low-temperature data from \citet{Ferguson2005} and data for the high-temperature, Compton-scattering dominated regime from \citet{Poutanen2017}. Electron conduction opacities are provided by \citet{Cassisi2007} and \citet{Blouin2020}. The OPAL Type II opacity tables, which consider varying C and O abundances, are utilized in our modelling of the He-burning phase. Nuclear reaction rates are sourced from JINA REACLIB \citep{Cyburt2010}, NACRE \citep{Angulo1999}, and additional tabulated weak reaction rates \citep{Fuller1985, Oda1994, Langanke2000}. Utilizing these tables, nuclear-burning networks that include the hot CNO cycles, the triple-alpha process, and C/O fusion, along with subsequent alpha captures and weak nuclear reactions, are employed in our modelling. Screening effects are included using the prescription by \citet{Chugunov2007}. Thermal neutrino loss rates are taken from \citet{Itoh1996}. The rates of energy loss from nuclear neutrinos, resulting from weak interactions and thermal neutrinos, are calculated by {\sc mesa}. 

The chemical composition utilized in this study is based on the \citet{Asplund_2009} solar mixture (A09). The initial mass fractions are assigned as follows: $X_{ini} = 0.738$ for H, $Y_{ini} = 0.248$ for He, and $Z_{ini} = 0.014$ for heavy elements. We adopt the Cox MLT description and set the mixing length parameter to a fixed value of $\alpha_{MLT} = 1.8$. These elements play a critical role in the core boundary evolution during helium burning. During the red giant branch phase, a Reimers' wind with $\eta$ = 0.1 is applied \citep{Reimers_1975}, whereas mass loss is neglected during the core hydrogen and core He-burning phases. When the input physics is taken into account appropriately, {\sc mesa} is capable of accurately modelling off-centre He flashes in the electron-degenerate cores of low-mass stars \citep{Paxton_2011, Bildsten_2012}. At all stages of the evolution, rotation is ignored in the models. Our evolutionary models include atomic diffusion, governed by temperature, chemical gradients, and gravitational settling. However, when applying the predictive mixing numerical model, atomic diffusion at our temporal resolution results in convective core splitting during He burning. As a result, we disable atomic diffusion in the predictive mixing case. Using a method given by \citet{Thoul_1994} and modified by \citet{Hu_2011} for non-Coulomb interactions, {\sc mesa} solves the Burgers flow equations \citep{Burgers_1969}. 

{\sc mesa} offers a time-step selection method considering changes in physical properties throughout the evolution of a star, whether absolute or relative. In this study, all evolutionary models for the He-burning phase utilize {\tt varcontrol\_target = 10$^{-6}$} to manage relative differences in the internal structure between consecutive stellar models. The upper limit for the time steps is set to {\tt max\_years\_for\_timestep = 25000 yr}. The Schwarzschild criterion for convective instability is applied during the evolutionary stages. The exponential diffusive overshoot description, based on \citet{Freytag_1996} and \cite{Herwig_2000}, facilitated mixing at all convective boundaries for all evolutionary stages except during helium burning. We increase the mesh density near the core boundary and around the Brunt-Väisälä maximum by using the parameter {\tt mesh\_delta\_coeff\_factor=0.05}. This ensures comprehensive coverage of the entire convective core, the semiconvective region, and the radiative zone above. During the He-burning phase, the size of the convective core can fluctuate significantly, either increasing or decreasing. To accurately capture these changing convective boundaries, our models need to have a sufficient number of mesh points around them. In the scenarios involving the penetration zone and the application of the CPM, we opt for a less dense mesh by setting {\tt mesh\_delta\_coeff\_factor} to 0.09. This choice is based on findings that higher mesh densities can introduce fluctuations during the second stage of helium burning, a phase where our study anticipates smooth growth in semiconvective regions. This problem is likely related to the splitting, merging and remeshing of cells inside and beyond the convective penetration zone.

\end{appendix}
\label{LastPage}

\end{document}